**Surface resistivity of hydrogenated amorphous carbon films:**

**Existence of intrinsic graphene on its surface**

S.S. Tinchev

Institute of Electronics, Bulgarian Academy of Sciences, Sofia 1784, Bulgaria

Abstract:

Surface resistivity of hydrogenated amorphous carbon films was measured as a function of the applied electrical field. The measured dependence shows a sharp ambipolar peak near zero gate voltage. Furthermore, we found that in some samples sheet resistance at the peak is as low as 7.5 k$\Omega$/sq. This value is the same order of magnitude as the sheet resistance of a defect free graphene monolayer. Therefore a conclusion is made that an intrinsic graphene with dimensions of at least millimeters exist on the surface of amorphous carbon films. These results can open new perspectives not only for graphene applications, but also for better understanding of this unique material.

## 1. Introduction

Graphene with its unique electrical properties attracts increasing interest for both fundamental science and applications. Today the most successfully graphene fabrication technology is CVD on metals followed by a transfer process on insulated substrates. Another method is fabrication of graphene from SiC by thermal decomposition which needs high temperatures to sublimate Si atoms and thus is incompatible with the existing silicon electronics. Recently [1] we proposed an idea for fabrication of graphene on the top of insulating amorphous carbon films by low-energy ion modification. The observed Raman spectra were typical for defected graphene - splitted D- and G-peaks and a broad 2D-peak. Back gated electrical measurements [2] show ambipolar electrical field effect typical for a single-layer graphene. However, during these experiments we observed similar effects also in the as deposited films, prior the ion modification. Here we present results of more carefully measurements of surface resistivity of hydrogenated amorphous carbon films.

## 2. Experimental

The films used in our experiments were 70 nm amorphous hydrogenated (a-C:H) carbon films deposited by DC PE CVD (plasma enhanced CVD) from benzene vapor diluted with argon. They are fabricated on the top of 300 nm thermal $SiO_2$ on Si. The $SiO_2$ serves also as an insulating layer, so a back-gate voltage can be applied to vary carrier concentration. These films can be produced with very different resistivity from soft graphitic-like low resistance films to hard, high resistance films by varying only the bias voltage in a DC PE CVD system.

Samples were measured with a lock-in amplifier and 4-terminals silver paint contacts in the AC version of Van der Pauw method of measurement. The referent voltage of the lock-in amplifier was transformed into 10 nA - 100 nA current and applied over the sample. The frequency of the AC current is 469 Hz. The voltage drop is detected and converted into sheet resistance by $R_{sq} = (\pi/\ln 2)R_{AB,CD}$. Here the resistance $R_{AB,CD}$ is given by $(V_{CD})/I_{AB}$, as usual in the Van der Pauw method of measurement. The current flows from contacts A to B and the voltage is measured across contacts C and D. Measurements were also made in 4-points in line configuration. In all cases the distances between contacts was about 1 mm.

The back-gate voltage from a digital – analog converter (swept from -10 V to +10 V) was applied to the back side of the silicon substrate through a silver paint contact. In this resistivity measurement the silicon substrate is acting as a gate and 300 nm $SiO_2$ is a gate insulator. Electrical resistance of the films was measured at room temperature and under ambient atmospheric conditions.

3. Results

The measured dependence of the sheet resistance of an as deposited sample on the gate voltage is shown in Figure 1. Obviously this curve is a superposition of a curve typical for a p-type semiconductor and a sharp ambipolar peak near zero gate voltage. To our opinion the sharp ambipolar peak near zero gate voltage is caused by the graphene surface layer and the superimposed curve typical for a p-type semiconductor which originates from the underlying amorphous carbon. The observed increase of the film resistivity typical for a p-type semiconductor as the gate voltage is strongly negative is consistent with movement of the Fermi level away from the valence band. Similarly, the decrease of the resistivity as the gate voltage is strongly positive is consistent with movement of the Fermi level closer to the valence band. The observed sharp peak near zero gate voltage resembles the ambipolar field effect in graphene because of its zero gap [3]. Positive gate voltage near the peak induces electrons, negative voltage induces holes and thus the resistance drops on both sides of the peak. The rapid decrease in resistance with adding charge carriers indicates high mobility of the carriers in the material. The peak is slightly shifted (+ 0.3 V) towards positive gate voltages, indicating a slight p-type doping. Adsorbed contamination on the sample surface could be a reason because resistivity measurement is carried out in air at ambient conditions and it is well known that being in essence a surface, graphene is extremely sensitive to contamination.

Our explanation that we observed ambipolar peak also in the as deposited films, prior the ion modification is that ion bombardment happens during the amorphous carbon film fabrication. The a-C:H films are usually deposited in low temperature plasmas from a hydrocarbon precursor gas, which is dissociated and ionized in the plasma. Accelerating radicals and ions bombarding onto the surface leads to film growth. Because the $sp^2$ is the more stable phase a $sp^2$-bonded material should be condensed on the surface. Indeed using TEM (transmission electron microscopy) and EELS (spatially resolved electron energy loss spectroscopy) $sp^2$-bonded layer about 1 nm thick was observed at the surface of the tetrahedral amorphous carbon films [4, 5]. The existence of the underlying $sp^3$-bonded material in the films is explained by the ion bombardment of the $sp^2$-bonded material, which is continuously converted into the $sp^3$-bonded material. Thus, it appears that the intrinsic $sp^2$-bonded layer is present on the surface of amorphous carbon films also without any post deposition ion modification. Evidence is measured electrical conductivity of diamond-like carbon as a function of film depth. It was found that the diamond-like carbon films are electrically nonhomogeneous in thickness [6].

There is also another possible mechanism which can explain the existence of intrinsic graphene of the surface of amorphous carbon. On the base of calculated surface energy and surface atom structure, it was found [7] that the surface of amorphous carbon will reconstruct into $sp^2$ sites. These conclusions are made for tetrahedral amorphous carbon thin films, but to our opinion this effect is more general and $sp^2$ surface layer intrinsic to the growth process should exist also in other types of amorphous carbon in particular in hydrogenated amorphous carbon films.

The existence of the $sp^2$-rich layer on the surface of tetrahedral (ta-C) amorphous carbon was early recognized as problem for fabrication of TFT (thin film) transistors from these materials. It has been shown by Davis et al [4, 5] that although bulk ta-C films have high $sp^3$ content, there is always a surface layer, rich in $sp^2$ bonding present. The graphitic layer prevents transistor action and has to be removed. In [8] this graphitic layer on the top surface of the ta-C film was etched away in oxygen plasma and ta-C thin film transistor was successfully fabricated.

In order to prove our suggestion that the measured curve in Fig. 1 consist of superimposed curve of the underlying amorphous carbon and the sharp ambipolar peak from graphen surface layer, we removed the surface layer by oxygen plasma. The etching was made in pulsed oxygen plasma for processing time of 10 s at voltage amplitude 740 V, pulse frequency of 66 kHz and pulse time of 10 μs. During the plasma etching the pressure of the chamber was $2.6 \times 10^{-1}$ Torr.

The result of the measurement of the surface resistance of the sample after oxygen plasma etching is shown in Figure 2. As expected the ambipolar peak disappeared and the measured dependence is typical curve for a p-type semiconductor. This is a proof that graphene surface layer was etched away from the film. This curve can be subtracted from the curve of the virgin film and thus the dependence of the removed surface layer can be found. Fig. 3 shows this dependence and now the ambipolar peak is more pronounced. The resistance peak is extremely narrow in comparison with graphene on $SiO_2$, which is much broader and shows a large offset from zero, typically some tens volts and more [3]. The small width of the resistivity peak at the charge neutrality point is an indication of a small carrier density fluctuation.

The sheet resistance at the peak maximum is about 7.5 kΩ /sq. This value is the same order of magnitude as the sheet resistance of a defect free graphene monolayer [3] and is very close to the value of 6.5 kΩ given by minimum quantum conductivity $4e^2/h$, where h is the Plank's constant, e is the charge of an electron. The sheet resistance can be converted to a bulk resistivity by multiplying by the thickness of the sample. Assuming thickness of the $sp^2$ bonded surface layer of about 1 nm as found from electron energy loss spectroscopy [4, 5] the estimated resistivity is about $10^{-7}$ Ω.cm, similar as the resistivity of the monolayer graphene.

In order to check if the resistivity has metallic or semiconductor type temperature dependence a measurement of the surface resistance was carried out at slightly higher temperature. Results of this measurement are plotted in Figure 4. They show that the background resistance decreased with increasing temperature whilst the ambipolar peak remained almost the same. This is also an additional indication for the existence of graphene on the surface of the as deposited amorphous carbon films.

Such ambipolar peak is observed also in high resistivity samples - Fig. 5. Here the surface resistivity measurement of this sample was repeated three times. Obviously one can recognize a general the p-type conductivity of the sample and fluctuating conductivity caused by a carriers hopping. Evidently is also that the hopping conductivity fluctuations are stronger for negative gate voltages because of smaller currier numbers in this region.

In a next experiment the sample was processed in argon plasma similar to the plasma during the film deposition. The processing time was 1 minute at pulsed unipolar voltage amplitude of 400 V at frequency of 66 kHz and pulse time of 10 µs. After the Ar-plasma treatment the ambipolar graphene peak is more pronounced - Fig. 6. This is also evidence that ion bombardment of the film surface is a cause for existence of intrinsic graphene of the film surface. Obviously is also that after the ion treatment the hopping effect is weaker and the sample is getting more resistive. This can be explained by further modification of the material and destroying the weaker links between individual graphene clusters.

In another sample (Figure 7) the gate voltage was swept in both directions: from – 10 V to + 10 V and back. There are some differences in the peak amplitude and their position. Nevertheless the ambipolar peak near zero gate voltage is clear visible. Surface resistance of all samples shows also strong photoconductivity - Figure 8. In these experiments the gate voltage was zero and the sample was simply illuminated and shadowed from a daily light in the room.

4. Conclusions

In conclusion, we report here on surface resistivity measurement of hydrogenated amorphous carbon films. It was found that the measured electrical field dependence shows a sharp ambipolar peak near zero gate voltage and minimum of the sheet resistance at the peak of 7.5 k$\Omega$ /sq. This value is the same order of magnitude as the sheet resistance of a defect free graphene monolayer. Because the distances between contacts in these measurements are about 1 mm one can make a conclusion that an intrinsic graphene exists at least over millimeter size areas on the surface of hydrogenated amorphous carbon films and probably on the surface of all other types of amorphous carbon films. This result can open new perspectives for carbon-based electronics and for better understanding of this unique material.

Figure captions:

Figure 1. The sheet resistance versus gate voltage of as deposited hydrogenated amorphous carbon films.

Figure 2. The sheet resistance vs. gate voltage of the film from Fig. 1 after oxygen plasma etching for 10 s.

Figure 3. The sheet resistance of the surface layer obtained by subtraction of Fig. 1 and Fig. 2.

Figure 4. The sheet resistance vs. gate voltage of as deposited hydrogenated amorphous carbon films for two temperatures: curve 1 at temperature of 22ºC, curve 2 at temperature of 50ºC.

Figure 5. The sheet resistance versus gate voltage of as deposited high-resistance sample of hydrogenated amorphous carbon films.

Figure 6. The sheet resistance vs. gate voltage of the film from Fig. 5 after argon plasma treatment for 1min.

Figure 7. The sheet resistance versus gate voltage if the gate voltage was swept in both directions.

Figure 8. Demonstration of strong photoconductivity of the surface resistance.

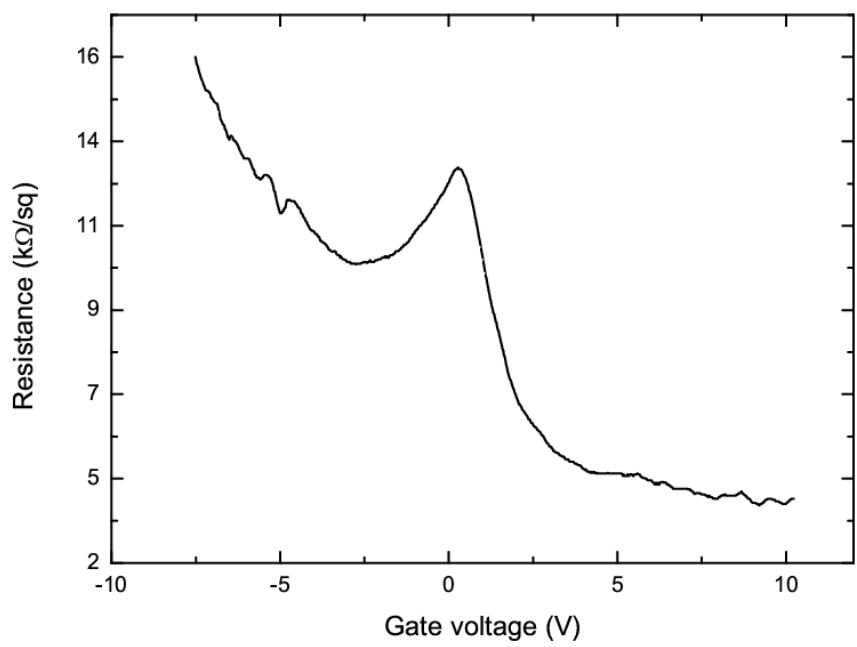

Figure 1.

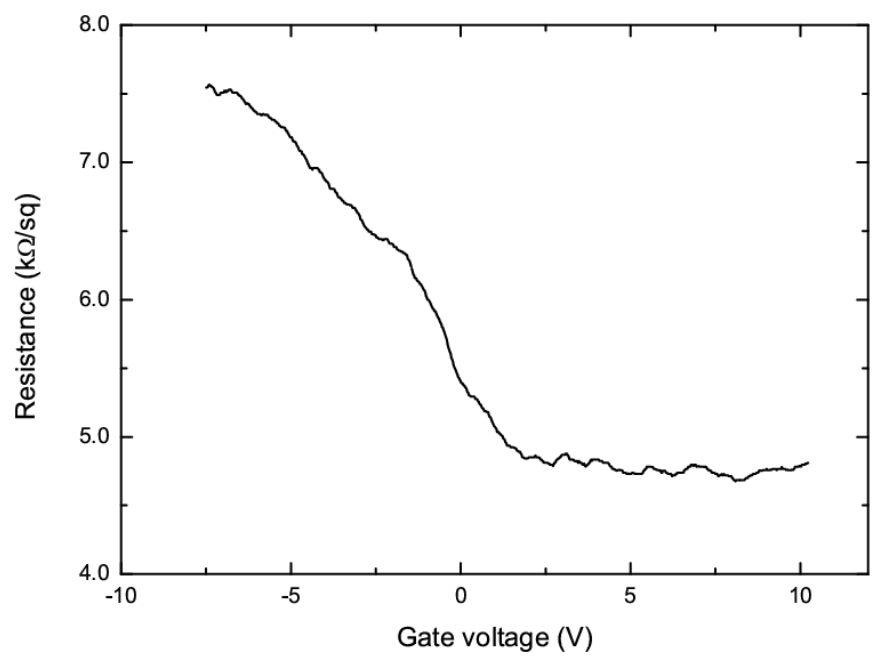

Figure 2.

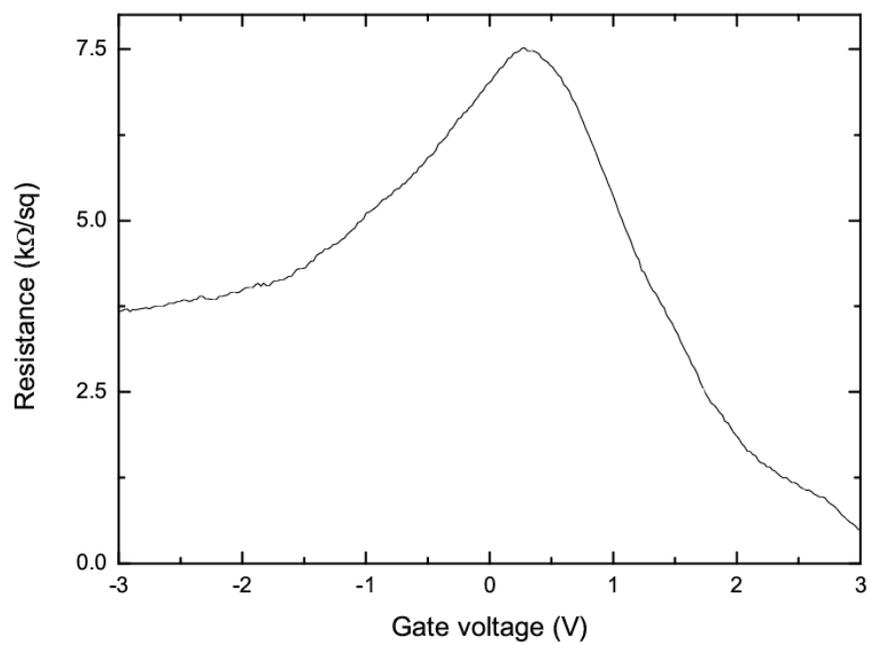

Figure 3.

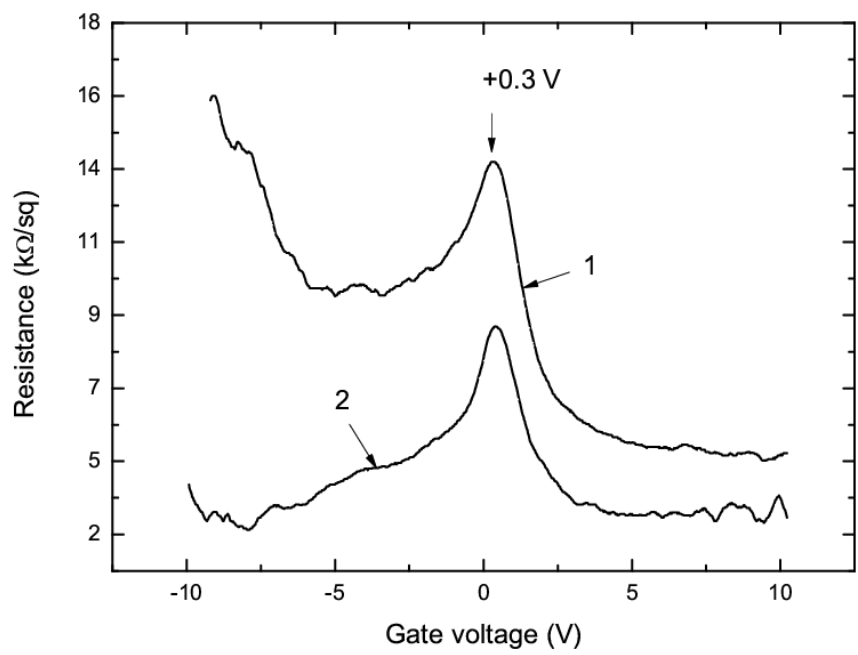

Figure 4.

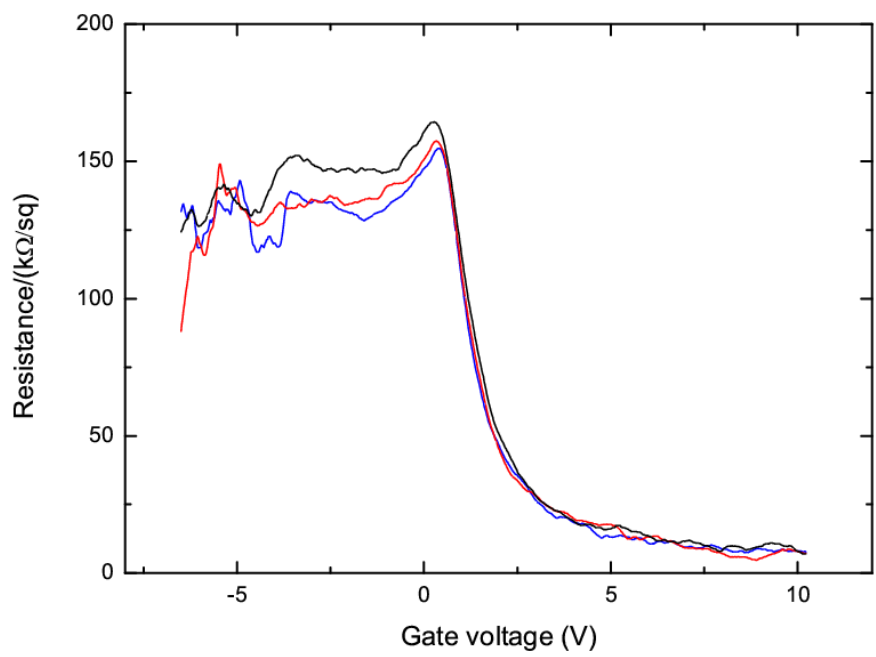

Figure 5.

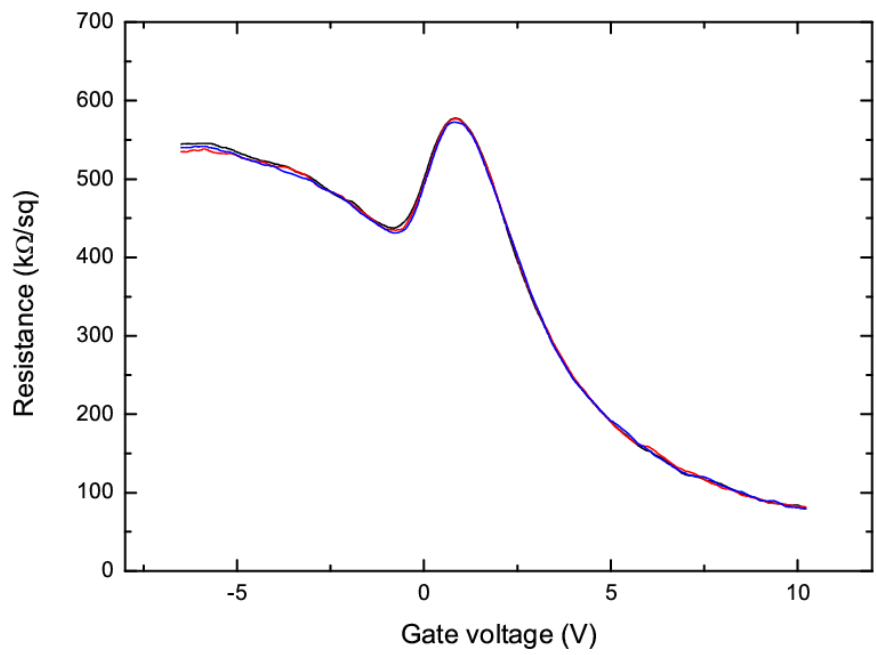

Figure 6.

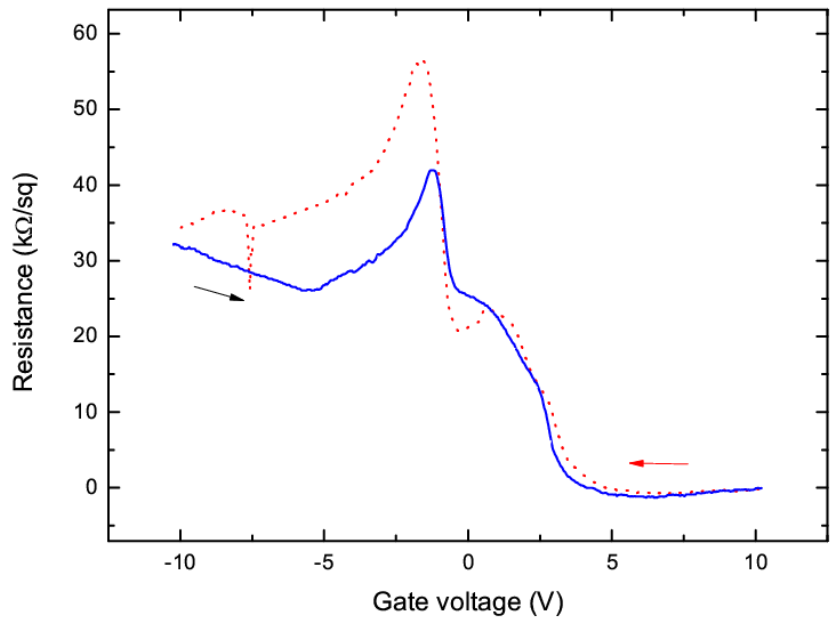

Figure 7.

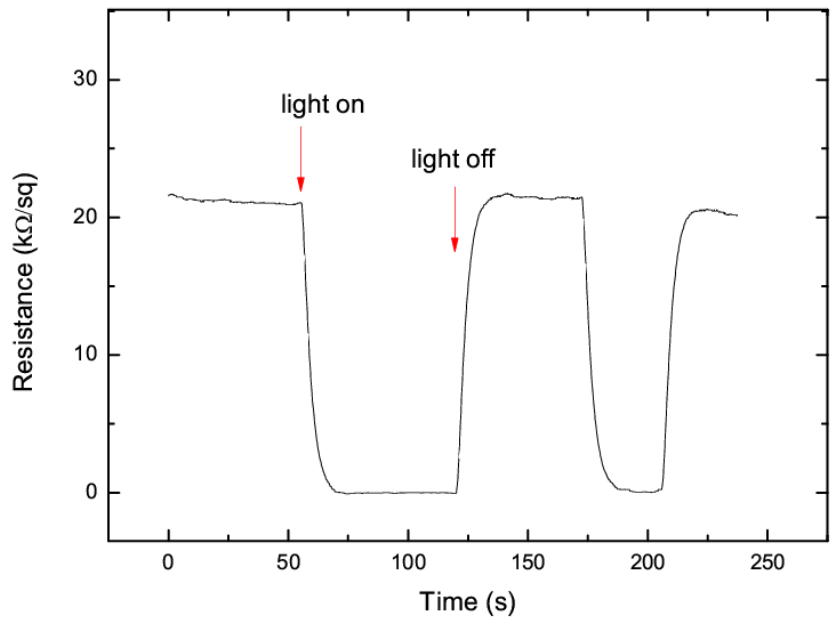

Figure 8.